# IMPROVING THE PERFORMANCE OF THE LINEAR SYSTEMS SOLVERS USING CUDA

**BOGDAN OANCEA[*]**
**TUDOREL ANDREI[**]**
**RALUCA MARIANA DRAGOESCU[***]**

**Abstract**
*Parallel computing can offer an enormous advantage regarding the performance for very large applications in almost any field: scientific computing, computer vision, databases, data mining, and economics. GPUs are high performance many-core processors that can obtain very high FLOP rates. Since the first idea of using GPU for general purpose computing, things have evolved and now there are several approaches to GPU programming: CUDA from NVIDIA and Stream from AMD. CUDA is now a popular programming model for general purpose computations on GPU for C/C++ programmers. A great number of applications were ported to CUDA programming model and they obtain speedups of orders of magnitude comparing to optimized CPU implementations. In this paper we present an implementation of a library for solving linear systems using the C-CUDA framework. We present the results of performance tests and show that using GPU one can obtain speedups of about of approximately 80 times comparing with a CPU implementation.*

## Introduction

Parallel computing can offer an enormous advantage regarding the performance for very large applications in almost any field: scientific computing, computer vision, databases, data mining, and economics. GPUs are high performance many-core processors that can obtain very high FLOP rates. Since the first idea of using GPU for general purpose computing, things have evolved and now there are several approaches to GPU programming: CUDA from NVIDIA and Stream from AMD. CUDA is now a popular programming model for general purpose computations on GPU for C/C++ programmers. A great number of applications were ported to CUDA programming model and they obtain speedups of orders of magnitude comparing to optimized CPU implementations.

Mark Harris[1]recognized for the first time the potential of using graphical processing units (GPU) for general purpose applications. Since then GPU programming models have evolved and there are several approaches to GPU programming now: CUDA (Compute Unified Device Architecture) from NVIDIA and APP (Stream) from AMD. A new standard OpenCL (Open Computing Language)[2] tries to unify different GPU general computing API implementations and to provide a general framework for writing programs executed across heterogeneous platforms consisting of both CPUs and GPUs.

In this paper will we use C-CUDA extension for developing a GPU accelerated library that implements direct and iterative methods for large linear systems. In our library we used the CUBLAS[3] library as a BLAS GPU accelerated library.

---

[*] Professor, Ph. D., "Nicolae Titulescu" University (email: bogdan.oancea@gmail.com).
[**] Professor, Ph. D., Bucharest Academy of Economic Studies (email: andreitudorel@yahoo.com).
[***] Assistant Lecturer, "Artifex" University (email: raluca_dragoescu@yahoo.com).

[1] Harris, Mark J., William V. Baxter III, Thorsten Scheuermann, and Anselmo Lastra.( 2003), Simulation of Cloud Dynamics on Graphics Hardware. In *Proceedings of the IGGRAPH/Eurographics Workshop on Graphics Hardware* 2003, pp. 92-101.

[2] Khronos OpenCL Working Group (2009), The OpenCL Specification - Version 1.0. The Khronos Group, Tech. Rep.

[3] NVIDIA (2007) CUDA – CUBLAS Library.

**Serial iterative and direct methods for solving large linear systems**

The classical approach for solving a linear system using iterative methods consists in Jacobi, Gauss-Seidel and SOR methods that are well known are presented in many textbooks[4].

For very large linear systems, the most appropriate iterative methods are the Krylov techniques[5]. Contrary to stationary iterative methods such as Jacobi or Gauss-Seidel, Krylov techniques use information that changes from iteration to iteration. For a linear system $Ax = b$, Krylov methods compute the $i^{th}$ iterate *x(i)* as :

$$x(i) = x(i-1) + d(i) \quad i = 1,2,...$$

Operations involved to find the $i^{th}$ update *d(i)* are only inner products, *saxpy* and matrix-vector products that has the complexity of $\Theta(n^2)$, so that Krylov methods are computational attractive comparing to the direct methods for linear systems that computes a decomposition of the matrix A into two triangular matrices.

Perhaps the best known and largely used in real applications Krylov method is the conjugate gradient method (CG). This method is used to solve symmetric positive definite (SPD) systems. The idea of the CG method is to update the iterates *x(i)* in such a manner to ensure the largest decrease of the objective function $\frac{1}{2}x'Ax - x'b$, while keeping the direction vectors *d(i)* *A*-orthogonal. This method can be implemented using only one matrix-vector multiplication per iteration. In exact arithmetic, the CG method gives the solution for at most *n* iterations. The complete description of the CG method can be found in (Golub, 1996).

Another Krylov method for general non symmetric systems is the Generalized Minimal Residuals (GMRES) introduced by (Saad, 1996). The pseudo-code for GMRES is:

**GMRES**

Given an initial solution x(0) compute r = b – Ax(0)
$\rho = ||r||_2$, $v(1) = r/\rho$, $\beta = \rho$
**for** k = 1,2,... **until** convergence
    **for** j = 1,2, ... k,
        h(j,k) = (Av(k))'v(j)
    **end**
    $v(k+1) = Av(k) - \sum_{j=1}^{k} h(j,k)v(j)$
    $h(k+1,k) = ||v(k+1)||_2$
    v(k+1,k) = v(k+1)/h(k+1,k)
**endfor**

The most difficult part of this algorithm is not to lose the orthogonality of the direction vectors *v(j).* To achieve this goal the GMRES method uses a Gram-Schmidt orthogonalization process. GMRES requires the storage and computation of an increasing amount of information, vectors *v* and matrix *H*. To overcome these difficulties, the method can be restarted after a chosen number of iterations *m*. The current intermediate results are used as a new starting point.

[4] Golub, G. H., and C. F. Van Loan, Matrix Computations (1996), Johns Hopkins Series in Mathematical Sciences, The Johns Hopkins University Press

[5] Saad, Y. (1996), Iterative Methods for Sparse Linear Systems, PWS Publishing Company.

Another Krylov method implemented by the authors is the BiConjugate Gradient method[6]. BiCG uses a different approach based upon generating two mutually orthogonal sequences of residual vectors and A-orthogonal sequences of direction vectors. The updates for residuals and for the direction vectors are similar to those of the CG method, but are performed using A and its transpose. The disadvantage of the BiCG method is an erratic behaviour of the norm of the residuals and potential breakdowns. An improved version, called BiConjugate Gradient Stabilized BiCGSTAB, is presented bellow:

**BiCGSTAB**

Given an initial solution x(0) compute r = b – Ax(0)

$\rho_0 = 1$, $\rho_1 = r(0)'r(0)$, $\alpha = 1$, $\acute{\omega} = 1$, p = 0, v = 0

**for** k = 1,2, ... **until** convergence

$\beta = (\rho_k / \rho_{k-1})(\alpha/\acute{\omega})$

$p = r + \beta(p - \acute{\omega} v)$

$v = Ap$

$\alpha = \rho_k/(r(0)'v)$

$s = r - \alpha v$

$t = As$

$\acute{\omega} = (t's)(t't)$

$x(k) = x(k-1) + \alpha p + \acute{\omega} s$

For the BiCGSTAB method we need to compute 6 *saxpy* operations, 4 inner products and 2 matrix-vector products per iteration and to store matrix A and 7 vectors of size *n*. The computational complexity of the method is $\Theta(n^2)$ like the other Krylov methods. The operation count per iteration cannot be used to directly compare the performance of BiCGSTAB with GMRES because GMRES converges in much less iterations than BiCGSTAB. We have implemented these iterative methods and run experiments to determine the possible advantages of them over the direct methods. The results of our experiments are presented in the next section.

The alternative to solve a linear system $Ax = b$ is the ***direct method*** that consists in two steps:

- First, the matrix A is factorized, $A = LU$ where *L* is a lower triangular matrix with 1s on the main diagonal and *U* is an upper triangular matrix; in the case of symmetric positive definite matrices, we have $A = LL^t$.
- Second, we have to solve two linear systems with triangular matrices: $Ly = b$ and $Ux = y$.

The standard LU factorization algorithm with partial pivoting is (Golub, 1996):

[6] Golub, G. H., and C. F. Van Loan, Matrix Computations (1996), Johns Hopkins Series in Mathematical Sciences, The Johns Hopkins University Press

**Right-looking LU factorization**
**for** k =1:n-1 **do**
    find ν with k≤ ν≤n such that $|A(\nu,k)| = \|A(k:n,k)\|_\infty$
    A(k,k:n)↔A(ν, k:n)
    p(k) = ν
    **if** A(k,k) ≠ 0 **then**
        A(k+1:n, k) = A(k+1:n,k)/A(k,k)
        A(k+1:n,k+1:n) = A(k+1:n,k+1:n) - A(k+1:n, k)
A(k, k+1:n)

The computational complexity of this algorithm is $\Theta(2n^3/2)$. After we obtain the matrix factors *L* and *U* we have to solve two triangular systems: $Ly = b$ and $Ux = y$. These systems are solved using forward and backward substitution that have a computational complexity of $\Theta(n^2)$, so the most important computational step is the matrix factorization. That's why we have to show a special attention to the algorithms for matrix factorization.

In practice, using actual computers with memory hierarchies, the above algorithm is not efficient because it uses only level 1 and level 2 BLAS operations[7]. As it is well-known, level 3 BLAS operations[8] have a better efficiency than level 1 or level 2 operations. The standard way to change a level 2 BLAS operations into a level 3 BLAS operation is delayed updating. In the case of the LU factorization algorithm we will replace *k* rank-1 updates with a single rank-*k* update.

We present a block algorithm for LU factorization that uses level 3 BLAS operations. The $n \times n$ matrix A is partitioned as in Figure 1. The $A_{00}$ block consists of the first *b* columns and rows of the matrix A.

$$\begin{bmatrix} A_{00} & A_{01} \\ A_{10} & A_{11} \end{bmatrix} = \begin{bmatrix} L_{00} & 0 \\ L_{10} & L_{11} \end{bmatrix} * \begin{bmatrix} U_{00} & U_{01} \\ 0 & U_{11} \end{bmatrix}$$

Figure 1. Block LU factorization

We can derive the following equations starting from A=LU:

$L_{00}U_{00} = A_{00}$ (1)

$L_{10}U_{00} = A_{10}$ (2)

$L_{00}U_{01} = A_{01}$ (3)

$L_{10}U_{01} + L_{11}U_{11} = A_{11}$ (4)

---

[7] Dongarra, J., J. Du Croz, S. Hammarling, and R. Hanson (1988): "An extended set of FORTRAN basic linear algebra subprograms", ACM Transactions on Mathematical Software, 14, (1), 1-17.

[8] Dongarra, J., J. Du Croz, S. Hammarling, and I. Duff (1990): "A set of level 3 basic linear Algebra subprograms", ACM Transactions on Mathematical Software, 16 (1), 1-17

Equations (1) and (2) perform the *LU* of the first *b* columns of the matrix *A*. Thus we obtain $L_{00}$, $L_{10}$ and $U_{00}$ and now we can solve the triangular system from equation (3) that gives $U_{01}$. The problem of computing $L_{11}$ and $U_{11}$ reduces to compute the factorization of the submatrix $A_{11}' = A_{11} - L_{10}U_{01}$ that can be done using the same algorithm but with $A_{11}'$ instead of A. The block LU factorization algorithm can now be derived easily: suppose we have divided the matrix A in column blocks with *b* columns in each block. The complete block LU factorization algorithm is given below.

**Block LU factorization**

```
for kb =1 to n-1 step b do
    bf = min(kb + b – 1, n)
    {LU factorization of A(kb : n, kb : bf ) with BLAS 2}
    for k = kb to bf do
        find k such that |A(k,i)| = ||A(i : n,i)||∞
        if i ≠ k then
            swap rows i and k
        endif
        A(i+1:n, i) = A(i+1:n, i)/A(i,i)
        A(i+1:n, i+1: bf) = A(i+1:n, i+1: bf ) - A(i+1:n, i) A(i, i+1: bf)
    endfor
    {Let L~ be unit lower triangular matrix b×b stored in A(kb : bf, kb : bf)}
    Solve triangular systems L~ Z = A(kb : bf, bf + 1 : n)
    Update A(kb : bf, bf + 1 : n) ← Z
    {Delayed updating}
    A(bf + 1 : n, bf + 1 : n) = A(bf + 1 : n, bf + 1 : n) − A(bf + 1 : n, kb : bf) A(kb : bf, bf + 1 : n)
endfor
```

The process of factorization is shown in Figure 2. The factorization of the current column block is done with the usual BLAS 2 operations and the active part of the matrix A will be updated with *b* rank-one updates simultaneously which in fact is a matrix-matrix multiplication (level 3 BLAS). If $n >> b$ almost all floating point operations are done in the matrix-matrix multiplication operation.

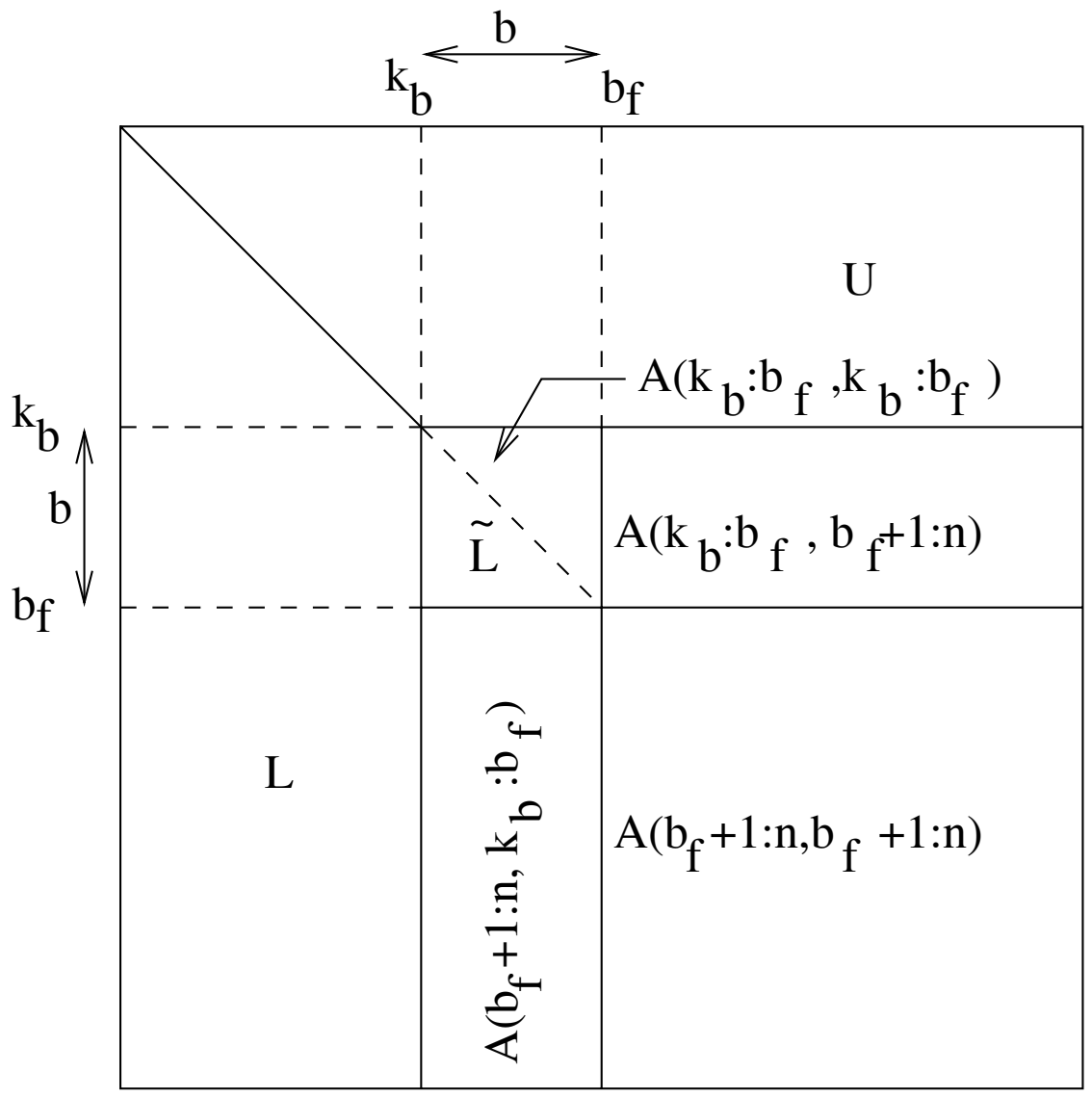

Figure 2. Block LU factorization with BLAS 3 operations

Cholesky factorization consists in finding the factorization of the form A = LL$^T$ where A is a symmetric positive definite (SPD) matrix. Figure 3 shows the partitioning of matrices A and L.

$$A = \begin{pmatrix} A_11 & * \\ A_21 & A_22 \end{pmatrix}$$

$$L = \begin{pmatrix} L_11 & 0 \\ L_21 & L_22 \end{pmatrix}$$

Figure 3. The partitioning of matrices A and L.

From $A = LL^T$ we can derive the following relations :

$A_{11} = L_{11}L_{11}^T$

$L_{21}L_{11}^T = A_{21}$

$A_{22} - L_{21}L_{21}^T = L_{22}L_{22}^T$

If matrix $L$ will overwrite the inferior triangle of $A$, then the Cholesky factorization consists in the following three computations:

$A_{11} \leftarrow L_{11} = Cholesky(A_{11})$

$A_{21} \leftarrow L_{21} = A_{21}L_{21}^{-T}$

$A_{22} \leftarrow A_{22} - L_{21}L_{21}^T$

**Parallel implementation of the direct and iterative algorithms using CUDA**

Our library implements LU and Cholesky factorization as directs methods and Jacobi, Gauss-Seidel, CG, GMRES and BiCGSTAB iterative methods.

The general flow of the solver implemented in our library is:

- Allocate memory for matrices and vectors in the host memory;
- Initialize matrices and vectors in the host memory;
- Allocate memory for matrices and vectors in the device memory;
- Copy matrices / vectors from host memory to device memory;
- Define the device grid layout:
  - Number of blocks
  - Threads per block
- Execute the kernel on the device;
- Copy back the results from device memory to host memory;
- Memory clean up.

We've used CUBLAS library in the implementation of the direct and iterative algorithms for performing BLAS operations. We also implemented the same algorithms in a single threaded program developed in C and run on CPU. The CPU library uses ATLAS[9] as a high performance BLAS implementation.

**Results**

We've tested our direct and iterative solvers for both single precision and double precision floating point numbers. For our tests we used a computer with Intel Core2 Quad Q6600 procesor running at 2.4 Ghz, 4 GB of RAM and a NVIDIA GeForce GTX 280 graphics processing unit (GPU) with 240 cores running at 1296 MHz, 1GB of video memory and 141.7 GB/sec memory bandwith. The operating system used was Windows Vista 64 bit.

We compared the results obtained using the CUDA code with the single threaded C implementation run on CPU. The CPU implementation of the direct and iterative algorithms used the optimized ATLAS library as a BLAS implementation. This gives better performances than a standard reference implementation of the BLAS.

Table 1 shows the speedup obtained by the C-CUDA implementation of the iterative solvers compared with the traditional CPU code for single precision floating point numbers and table 2 shows the speedup for double precision numbers. From the results presented below one can see that GPU outperforms CPU for numerical computations.

Comparing the results for each method, it can be noticed that BiCGSTAB has better performances than the other methods. For GMRES, in our experiments we restarted the method after 35 iterations. The tolerance for the solution was fixed at $10^{-4}$ for all methods. For our experiments we have considered linear systems containing between 2000 and 20000 variables.

Table 3 shows the speedup of the CUDA implementation of the direct method for linear systems compared with a single threaded C implementation (the standard block-level implementation that can be found in LAPACK). We considered linear systems with 500 to 3500 equations.

Our performance results show the net advantage of GPU computing compared to the classical CPU code. The results also emphasize the advantage of the iterative solutions compared with the direct solution. Another advantage of using CUDA programming model is that the code can be easier

[9] Whaley, R. C., A. Petitet, and J. Dongarra (2001), "Automated Empirical Optimization of Software and the ATLAS project", Parallel Computing, 27(1-2), 3-35

read and support. The major drawback of CUDA is that it is only available for NVIDIA devices. A port of our library to OpenCL is intended for the future.

Table 1. Speedup of the CUDA library for single precision FP

| Matrix dimension | Speedup | | | |
|---|---|---|---|---|
| | Jacobi | Gauss-Seidel | GMRES(35) | BiCGSTAB |
| 2000 | 67.4 | 69.3 | 78.3 | 82.2 |
| 4000 | 56.2 | 65.5 | 81.8 | 84.5 |
| 8000 | 68.3 | 67.4 | 80.1 | 81.9 |
| 12000 | 66.7 | 68.4 | 81.4 | 84.1 |
| 16000 | 71.1 | 69.2 | 79.3 | 86.0 |
| 20000 | 72.8 | 69.9 | 81.3 | 86.9 |

Table 2. Speed up for double precision FP

| Matrix dimension | Speedup | | | |
|---|---|---|---|---|
| | Jacobi | Gauss-Seidel | GMRES(35) | BiCGSTAB |
| 2000 | 35.2 | 36.1 | 39.6 | 41.7 |
| 4000 | 36.1 | 36.0 | 41.2 | 42.3 |
| 8000 | 29.1 | 35.2 | 41.6 | 43.6 |
| 12000 | 33.6 | 37.8 | 40.5 | 43.9 |
| 16000 | 32.3 | 35.9 | 42.8 | 44.0 |
| 20000 | 35.6 | 37.1 | 43.2 | 46.1 |

Table 3. The speedup of the direct method based on LU factorization for double precision

| Matrix dimension | C-CUDA |
|---|---|
| 500 | 8.99 |
| 1000 | 12.45 |
| 1500 | 11.41 |
| 2000 | 16.78 |
| 2500 | 16.23 |
| 3000 | 14.39 |

Table 4. The speedup of the direct method based on Cholesky factorization for SPD matrices double precision

| Matrix dimension | C-CUDA |
|---|---|
| 500 | 13.50 |
| 1000 | 19.75 |
| 1500 | 19.71 |
| 2000 | 23.17 |
| 2500 | 24.45 |
| 3000 | 22.585 |
| 3500 | 23.90 |

## Conclusions

We developed a C-CUDA library that implements the direct method with LU and Cholesky factorization and Jacobi, Gauss-Seidel and non-stationary iterative methods (GMRES, BiCGSTAB). The matrix-vector and matrix-matrix computations were done using CUBLAS routines. We compared the performance of our CUDA implementation with classic programs written to be run on CPU. Our performance tests show speedups of approximately 80 times for single precision floating point numbers and 40 times for double precision for the iterative methods and about 10-25 for the direct method with double precision FP. The lower figures of the speedups for direct methods may come from the memory bandwidth.

These results show the immense potential of the GPU accelerated numerical computations. In the future we intend to extend our direct and iterative solver library and to port it to OpenCL.

## References

- AMD (2008), ATI Stream Computing - Technical Overview. AMD, Tech. Rep.
- AMD (2011), *AMD Accelerated Parallel Processing – Programming Guide*.
- Agullo, E., Augonnet, C., Dongarra, J., Faverge, M., Langou, J., Ltaief, H., Tomov, S. (2011), *LU Factorization for Accelerator-based Systems*, IEEE/ACS AICCSA 2011, Sharm-El-Sheikh, Egypt, December, 2011
- Barron, Iann M. (1978), D. Aspinall. ed. "The Transputer". The Microprocessor and its Application: an Advanced Course, Cambridge University Press.
- K. E. Batcher, (1980), Design of a Massively Parallel Processor, *IEEE Transactions on Computers*, Vol. C29, September, pp. 836–840.
- Dongarra, J., J. Du Croz, S. Hammarling, and I. Duff (1990): "A set of level 3 basic linear Algebra subprograms", ACM Transactions on Mathematical Software, 16 (1), 1-17.
- Dongarra, J., J. Du Croz, S. Hammarling, and R. Hanson (1988): "An extended set of FORTRAN basic linear algebra subprograms", ACM Transactions on Mathematical Software, 14, (1), 1-17.
- Doornik, J. A., D. F. Hendry, and N. Shephard (2007): "Parallel Computation in Econometrics: A Simplified Approach" Chapter 15 in Handbook of Parallel Computing and Statistics, Chapman & Hall/CRC, 449-476.
- Golub, G. H., and C. F. Van Loan, Matrix Computations (1996), Johns Hopkins Series in Mathematical Sciences, The Johns Hopkins University Press.
- Allan Gottlieb, Ralph Grishman, Clyde P. Kruskal, Kevin P. McAuliffe, Larry Rudolph, Marc Snir, (1982), *The NYU Ultracomputer—designing a MIMD, shared-memory parallel machine*, ISCA '82 Proceedings of the 9th annual symposium on Computer Architecture, pp. 27 – 42.
- Harris, Mark J., William V. Baxter III, Thorsten Scheuermann, and Anselmo Lastra.( 2003), Simulation of Cloud Dynamics on Graphics Hardware. In *Proceedings of the IGGRAPH/Eurographics Workshop on Graphics Hardware* 2003, pp. 92-101.
- INTEL (2012), Microprocessor Quick reference guide.
- Khronos OpenCL Working Group (2009), The OpenCL Specification - Version 1.0. The Khronos Group, Tech. Rep.
- Nath, R., Tomov, S., Dong, T., Dongarra, J. (2011), *Optimizing Symmetric Dense Matrix-Vector Multiplication on GPUs*, ACM/IEEE Conference on Supercomputing (SC'11), Seattle, WA, November 12-18, 2011.
- NVIDIA, (2010), CUDA C Programming Guide, Version 4.0.
- NVIDIA (2007) CUDA – CUBLAS Library.
- NVIDIA, (2011), NVIDIA CUDA C Programming Guide, version 4.0.
- Oancea, Bogdan, Ion. Gh. Rosca, Tudorel Andrei, Andreea Iluzia Iacob,(2011), Evaluating Java performance for linear algebra numerical computations, Procedia Computer Science, vol. 3, pp. 474-478.
- Saad, Y. (1996), Iterative Methods for Sparse Linear Systems, PWS Publishing Company.
- Sam Stone, Justin P. Haldar, Stephanie C. Tsao, Wen-mei W. Hwu, Zhi-Pei Liang, Bradley P. Sutton, (2008), *Accelerating Advanced MRI Reconstructions on GPUs,* Proceedings of the 5th International Conference on Computing Frontiers, May 5-7, http://doi.acm.org/10.1145/1366230.1366274.

- Lewis W. Tucker, George G. Robertson, (1988), Architecture and Applications of the Connection Machine, *Computer*, vol. 21, no. 8, pp. 26–38.
- Whaley, R. C., A. Petitet, and J. Dongarra (2001), “Automated Empirical Optimization of Software and the ATLAS project”, Parallel Computing, 27(1-2), 3-35.